\documentclass[12pt]{article}

\usepackage{amsmath,amsfonts,amssymb}

\newcounter{tabl}
\newcommand{\be}{\begin{equation}}
\newcommand{\ee}{\end{equation}}
\newcommand{\beq}{\begin{eqnarray}}
\newcommand{\eeq}{\end{eqnarray}}
\newcommand{\bea}[2]{\be\label{#2}\begin{array}{#1}}
\newcommand{\eea}{\end{array}\ee}

\def\({\left(}
\def\){\right)}
\def\[{\left[}
\def\]{\right]}

\def\11{1\!\! 1}

\def\hf{{1\over 2}}

\def\eps{\varepsilon}

\def\vt{\vartheta}

\def\vrh{\varrho}

   \def\CD {{\cal D}}

\def\bxi{\overline{\xi}}
\def\bzeta{\overline{\zeta}}
\newcommand{\tom}{{\tilde \omega}}

\begin{document}
%
%

\title{Immirzi parameter and \\ fermions with non-minimal coupling}

\author{Sergei Alexandrov\thanks{email: Sergey.Alexandrov@lpta.univ-montp2.fr}
}

\date{}

\maketitle

\vspace{-1cm}

\begin{center}
\emph{Laboratoire de Physique Th\'eorique \& Astroparticules} \\
\emph{Universit\'e Montpellier II, 34095 Montpellier Cedex 05, France}


\end{center}

\vspace{0.1cm}

\begin{abstract}

We clarify the role played by the Immirzi parameter in classical gravity coupled to fermions.
Considering the general non-minimal coupling, we show that, although the torsion
depends explicitly on the Immirzi parameter, in a suitable parametrization
the effective action obtained by integrating out the spin-connection is independent of it.
Thus the Immirzi parameter is not detectable in classical theory even after coupling of fermions.

\end{abstract}





\vspace{0.6cm}

Recently, coupling of fermions to classical general relativity has attracted much attention.
The interest was initiated by the work \cite{Perez:2005pm} where it has been noted that
the minimal coupling to fermions makes gravity sensitive to the so called Immirzi parameter \cite{imir}.
This parameter, which we call $\beta$, appears
through the action generalizing the standard Hilbert--Palatini formulation \cite{Holst:1995pc}
\be
S_{G}[e,\omega]=\frac{1}{16\pi G} \int d^4x\, e\,
e_I^\mu e_J^\nu
\(R^{IJ}_{\mu\nu}(\omega)-
\frac{1}{\beta}\star R^{IJ}_{\mu\nu}(\omega)\),
\label{palat}
\ee
where $e^I_\mu$ is the tetrad field, $R^{IJ}_{\mu\nu}(\omega)$ is
the curvature of the spin-connection $\omega^{IJ}_\mu$, and
the star operator is the Hodge operator defined as
$\star \omega^{IJ}_\mu=\frac 12 {\eps^{IJ}}_{KL}\omega^{KL}_\mu $.
In pure gravity $\beta$ does not affect the equations of motion.
However, in \cite{Perez:2005pm} it was demonstrated that once the minimally coupled action
\be
S_{ F\rm -min}[e,\omega,\psi]
=\frac{i}{2}\int d^4x\, e\, e_I^\mu\(\overline{ \psi} \gamma^I \CD_\mu \psi-\overline{ \CD_\mu\psi} \gamma^I \psi\)
\label{minferm}
\ee
is added and the spin-connection is integrated out, one obtains the following effective
4-fermion interaction term with a $\beta$-dependent coupling constant
\be
S_{\rm int-min}[e,\psi]=-\frac{3}{2}\,\pi G\, \frac{\beta^2}{1+\beta^2}\int d^4 x\, e\,A^I A_I,
\label{minint}
\ee
where $A^I=\overline{\psi} \gamma^5 \gamma^I\psi$ is the axial current. Thus, it was concluded
that measuring the strength of such 4-fermion interaction can provide an information
about the Immirzi parameter. In other words, $\beta$ was argued to be in principle an observable parameter.
Later this result was generalized to a non-minimal coupling which was shown to lead to parity violation effects
\cite{Freidel:2005sn,Khriplovich:2005jh,Randono:2005up}.

On the other hand, in \cite{Mercuri:2006um} it was argued that the minimal coupling is inconsistent
because it leads to a torsion represented as a sum of a vector and a pseudovector. At the same time,
a particular non-minimal coupling has been suggested which cancels the effect of the Immirzi parameter
so that the latter drops out from the theory as in pure gravity.
However, the inconsistency argument is not convincing. The fact that the torsion does not have
a definite transformation property under parity changing simply means that
the theory is not parity invariant which might well be the case.

This collection of results gave rise to confusing statements in the literature
about the fermion couplings and the role of the Immirzi parameter in the classical theory.
This issue, being important by itself, becomes especially meaningful regarding the status of
the Immirzi parameter in quantum theory.
The standard results of loop quantum gravity (LQG) approach show that it enters the spectra
of geometrical operators as a scaling factor thus becoming an observable parameter
\cite{Rovbook,Ashtekar:1996eg,Ashtekar:1997fb}.\footnote{An attempt to consistently include the minimally coupled
fermions in this framework can be found in \cite{Thiemann:1997rq,Bojowald:2007nu}.}
On the other hand, the approach called {\it covariant} loop quantum gravity
suggests a quantization with results independent of $\beta$ and argues that the standard LQG
is anomalous \cite{SA,AV,SAcon,AlLiv}.

In this letter we would like to clarify the situation at the classical level.
For this, we suggest to consider the general {\it non-minimally} coupled action quadratic in fermions
\be
S_{F}[e,\omega,\psi]=\frac{i}{2}\int d^4x\, e\, e_I^\mu\(\overline{ \psi} \gamma^I
\(\zeta-i\xi\gamma^5 \)\CD_\mu \psi-\overline{ \CD_\mu\psi} \(\bzeta-i\bxi\gamma^5 \)\gamma^I \psi\),
\label{nonminferm}
\ee
where $\zeta$ and $\xi$ are two complex parameters.
We will work with their real and imaginary
parts which we denote as
\be
\zeta=\eta+i\theta,
\qquad
\xi=\rho+i\tau.
\ee
Actually, one of these four parameters can be absorbed by rescaling the fermion field $\psi$,
but we leave all of them to have more symmetric equations.

The action \eqref{nonminferm} is a natural combination of the actions considered in
\cite{Freidel:2005sn} and \cite{Mercuri:2006um}. One reproduces them for $\xi=0$ and $\theta=\tau=0$,
respectively. As one can check, the results we are going to present reduce to those
of \cite{Freidel:2005sn,Mercuri:2006um} in these particular limits.

Our primary goal is to solve the equations of motions with respect to the spin-connections and,
substituting the result into the initial action, to find an effective theory for the tetrad
coupled to the fermions.
It is convenient to represent the spin-connection as the sum of the torsion-free connection $\tom_\mu^{IJ}$
and the contorsion tensor $C_\mu^{IJ}$ which is closely related to the torsion. The torsion-free connection
satisfies
$\tilde\CD_{[\mu}e_{\nu]}^I=0$ and therefore can be expressed in terms of the tetrad so that one has
$\omega_\mu^{IJ}=\tom_\mu^{IJ}(e)+C_\mu^{IJ}$.
Then it is not difficult to obtain the following result for the contorsion
tensor\footnote{We use the conventions of \cite{Freidel:2005sn}. Note that there is a sign difference
with respect to the definition of $\beta$ used in \cite{Mercuri:2006um}.}
\beq
e^\mu_K C_\mu^{IJ}&=&-4\pi G\, \frac{\beta^2}{1+\beta^2}\[
\(\rho+\frac{\eta}{\beta}\)\delta_K^{[I} A^{J]}-\hf\(\eta-\frac{\rho}{\beta}\){\eps^{IJ}}_{KL} A^{L}
\right.
\nonumber \\
&& \left. \qquad
+\(\theta-\frac{\tau}{\beta}\)\delta_K^{[I} V^{J]}+\hf\(\tau+\frac{\theta}{\beta}\){\eps^{IJ}}_{KL} V^{L}
\],
\label{torsion}
\eeq
where $V^I=\overline{\psi} \gamma^I\psi$ is the vector current and the axial current $A^I$ was defined above.
After substitution of \eqref{torsion} into the action $S_G+S_F$, one finds
\beq
S_{\rm eff}[e,\psi]&=&\frac{1}{16\pi G} \int d^4x\, e\,  e_I^\mu e_J^\nu
R^{IJ}_{\mu\nu}(\tom)+
i\int d^4x\, e\, e_I^\mu \overline{ \psi} \(\eta- \tau\gamma^5\)\gamma^I\tilde\CD_\mu \psi
+S_{\rm int}[e,\psi],
\label{action}
\eeq
where the interaction term reads
\beq
S_{\rm int}[e,\psi]&=&-\frac{3}{2}\, \pi G \, \frac{\beta^2}{1+\beta^2}\int d^4 x\, e\[
\(\eta^2-\frac{2\eta\rho}{\beta}-\rho^2\)A^2+\(\tau^2+\frac{2\tau\theta}{\beta}-\theta^2\)V^2
\right. \nonumber \\
&& \left. \qquad
-\(\eta\tau+\theta\rho+\frac{\eta\theta-\tau\rho}{\beta}\)A_I V^I\].
\label{effint}
\eeq

The effective interaction \eqref{effint}
seems to explicitly depend on the Immirzi parameter. However, this does not allow us to make any predictions
concerning $\beta$ since a simple redefinition of the parameters removes the dependence.
Indeed, let us redefine two of the parameters $\theta$ and $\rho$ of the non-minimal coupling
in terms of new parameters $\vt$ and $\vrh$ as follows
\be
\theta=\sqrt{1+\frac{1}{\beta^2}}\,\vt+\frac{\tau}{\beta},
\qquad
\rho=\sqrt{1+\frac{1}{\beta^2}}\,\vrh-\frac{\eta}{\beta}.
\label{paramet}
\ee
Then both, the contorsion tensor and the interaction term,
depend only on the following combinations of currents
\be
J=\eta A-\tau V,
\qquad
Y=\vrh A+ \vt V
\ee
and are given by very simple expressions
\beq
&
\displaystyle e^\mu_K C_\mu^{IJ}=4\pi G \[\hf\, {\eps^{IJ}}_{KL}J^L
-\frac{\beta}{\sqrt{1+\beta^2}}\(\delta_K^{[I} Y^{J}+\frac{1}{2\beta}\, {\eps^{IJ}}_{KL}Y^L\)\],
&
\label{torsnew}
\\
&
\displaystyle S_{\rm int}[e,\psi]=-\frac{3}{2}\, \pi G \int d^4 x\, e\[ J^2-Y^2\].
&
\label{effintnew}
\eeq

The result \eqref{effintnew} together with \eqref{action} demonstrates that
in the parametrization \eqref{paramet} the effective dynamics of fermions coupled to the metric
does not depend on the Immirzi parameter which completely disappears from $S_{\rm eff}$.
The dependence on $\beta$ was absorbed into the new couplings $\vt$ and $\vrh$ which
together with $\eta$ and $\tau$ are the only measurable quantities at this
level.\footnote{Notice that the introduction of the currents $J$ and $Y$, which turn out to contain
all dependence of the couplings, is not essential for the main result. In fact, the parity transformation
does not preserve the form of these currents
and therefore they do not seem to have a direct physical interpretation.}

For non-vanishing $\vrh$ and $\vt$, the torsion however still carries a dependence of $\beta$.
But we do not have a direct access to it because the torsion manifest itself only through effective
phenomena similar to the four-fermion interaction we have found here \cite{Shapiro:2001rz}.
Thus, we conclude that in this framework the Immirzi parameter remains unmeasurable as in pure gravity.

In the particular case $\vrh=\vt=0$ the current $Y$ vanishes and
there are no second terms, both in the contorsion tensor \eqref{torsnew} and in the effective
interaction \eqref{effintnew}. In particular, this ensures that
$\CD_{\mu}(e\, e^\mu_I)=e \eta_{IK} e^\mu_J C_\mu^{JK}=0$ and the additional term in the action
\eqref{palat} introducing $\beta$ coincides with the Nieh-Yan invariant \cite{Mercuri:2006um}.
Taking this into account, it seems that in the presence of $\beta$ the choice $\vrh=\vt=0$
is physically distinguished and in some sense it plays the role of the minimal coupling of the
usual Einstein--Cartan theory.

\

{\bf Acknowledgements:} The author is grateful to Philippe Roche for discussions and to
Carlo Rovelli for correspondence.
This research is supported by CNRS and by the contract ANR-06-BLAN-0050.

\end{document}